\documentstyle[prl,aps,amsfonts,amssymb,floats,epsfig]{revtex}

\begin{document}

\draft
\twocolumn[\hsize\textwidth\columnwidth\hsize\csname @twocolumnfalse\endcsname

\title{Charge-order-induced sharp Raman peak in Sr$_{14}$Cu$_{24}$O$_{41}$}

\author{K.P. Schmidt$^1$, C. Knetter$^1$, M. Gr\"{u}ninger$^2$, and G.S.~Uhrig$^1$}
\address{$^1$Institut f\"{u}r Theoretische Physik, Universit\"{a}t zu
  K\"{o}ln, Z\"{u}lpicher Stra{\ss}e 77, D-50937 K\"{o}ln, Germany}
\address{$^2$II. Physikalisches Institut, Universit\"{a}t zu
  K\"{o}ln, Z\"{u}lpicher Stra{\ss}e 77, D-50937 K\"{o}ln, Germany}

\date{revised version: February 3, 2003}

\maketitle
\begin{abstract}
In the two-leg $S$=1/2 ladders of Sr$_{14}$Cu$_{24}$O$_{41}$ a modulation of
the exchange coupling arises from the charge order within the other structural
element, the CuO$_2$ chains.
In general, breaking translational invariance by modulation causes gaps within
the dispersion of elementary excitations.
We show that the gap induced by the charge order can drastically change the
magnetic Raman spectrum leading to the sharp peak observed in
Sr$_{14}$Cu$_{24}$O$_{41}$. This sharp Raman line gives insight in the
charge-order periodicity and hence in the distribution of carriers.
The much broader spectrum of La$_6$Ca$_8$Cu$_{24}$O$_{41}$ reflects
the response of an undoped ladder in the absence of charge order.
\end{abstract}


%75.40.Gb Dynamic properties (dynamic susceptibility, spin waves,
% spin diffusion, dynamic scaling, etc.)
%75.10.Jm Quantized spin models
%75.50.Ee Antiferromagnetics
%74.25.Ha Magnetic properties in superconductivity

\vskip2pc]
%]
\narrowtext

Understanding the complex interplay of spin and charge degrees of freedom in
doped quantum-spin systems is a key issue in condensed matter physics.
This interplay governs in particular the physics of the planar
high-$T_c$ superconducting cuprates. In the telephone-number compounds
A$_{14}$Cu$_{24}$O$_{41}$ it gives rise to a variety of interesting ground
states. In La$_6$Ca$_8$Cu$_{24}$O$_{41}$ the ladders form an insulating spin
liquid \cite{carte96}.
Sr$_{14-x}$Ca$_{x}$Cu$_{24}$O$_{41}$ becomes superconducting under external
pressure for $x \! \gtrapprox \! 11.5$ \cite{uehar96,nagat98}
whereas an insulating charge-ordered state is favored for
$x \! \lessapprox \!  5$
\cite{carte96,regna99b,matsu99,fukud02,brade02,ammer00,katae01},
although the average copper valence does not depend on $x$. The different
properties are usually attributed to the different distribution of charges
between the two subsystems \cite{nucke00,mizun97}, Cu$_2$O$_3$ two-leg ladders
and CuO$_2$ chains \cite{mccar88}. 
But the interesting physics linked to the differing periodicity of ladders and
chains  \cite{frost97b} is usually neglected.

Raman scattering offers a powerful tool to examine the spectral
density of magnetic excitations and thus provides
important information on the kinetics and on the interactions of the elementary
excitations. In the undoped spin liquid La$_6$Ca$_8$Cu$_{24}$O$_{41}$,
the ladders
show a very broad two-triplet Raman line with slightly different peak positions
for leg-leg and rung-rung polarization \cite{sugai99}. This agrees very well
 with
theoretical results \cite{schmi01}. A very different line shape, however, is
 found
in charge-ordered Sr$_{14}$Cu$_{24}$O$_{41}$,
 which shows a peculiar sharp peak that
is observed at the same frequency in both polarizations \cite{sugai99,gozar01}.
This sharp response poses a challenge to the understanding of the
cuprate ladders
and offers the opportunity to study the interplay of spin and charge degrees of
freedom in this fascinating system.

The sharpness of the Raman peak in Sr$_{14}$Cu$_{24}$O$_{41}$
is in strong contrast to the very broad two-magnon Raman line observed in the
 undoped
two-dimensional (2D) high-$T_c$ cuprates, which is
still the subject of  controversial discussions. Gozar {\em et al.}
\cite{gozar01}
argued that the observation of  a very sharp two-triplet Raman line in a
one-dimensional
(1D) spin liquid suggests that the large width found in 2D cannot be
attributed to quantum
fluctuations.

Here, we challenge this point of view by providing a clear explanation for the
  Raman
data of Sr$_{14}$Cu$_{24}$O$_{41}$. We propose that the charge-order
superstructure gives
rise to a modulation of the exchange coupling along the ladders.
The concomitant backfolding of the dispersion of the elementary triplet opens
 gaps at the crossing  points. This in turn can have
 a drastic effect on the Raman line shape,
which we calculate using continuous unitary transformations (CUTs)
\cite{knett00a,schmi01}.
The high resolution accessible by the CUT approach is decisive to account for
 the very
narrow peak we are aiming at. Our results with and without charge order
excellently describe
the Raman data of Sr$_{14}$Cu$_{24}$O$_{41}$ and
 La$_{6}$Ca$_{8}$Cu$_{24}$O$_{41}$, respectively.

For zero hole doping, the minimal model for the magnetic properties of the
$S=1/2$
two-leg ladders in A$_{14}$Cu$_{24}$O$_{41}$ is an antiferromagnetic
Heisenberg Hamiltonian plus a cyclic four-spin exchange term
$H_{\rm cyc}$ \cite{matsu00a,matsu00b,nunne02}
\begin{mathletters}
\label{eq:Hamiltonian}
\begin{eqnarray}
\label{hamil1}
&&H = J_\perp \sum\limits_i {\bf S}_{i,1}
{\bf S}_{i,2} + J_\parallel \sum_{i,\tau} {\bf S}_{i,\tau} {\bf S}_{i+1,\tau}
+H_{\rm cyc}\\
\label{hamil2}
&&H_{\rm cyc}= J_{\rm cyc}\sum_i K_{(i,1),(i,2),(i+\!1,2),(i+\!1,1)}
 \\
&&K_{1234} = \\\nonumber
&&\qquad({\bf S}_1 {\bf S}_2)({\bf S}_3 {\bf S}_4) + ({\bf
S}_1 {\bf S}_4)({\bf S}_2 {\bf S}_3) - ({\bf S}_1 {\bf S}_3)({\bf S}_2 {\bf
S}_4)\quad
\end{eqnarray}
\end{mathletters}
where $i$ denotes the rungs and $\tau\in\{1,2\}$ the legs.
The exchange couplings along the rungs and along the legs are denoted by
$J_\perp$
and $J_\parallel$, resp.. There is also another way to include the
 leading
four-spin exchange term by cyclic permutations \cite{brehm99,nunne02} which
differs in certain two-spin terms from Eq.\ (\ref{eq:Hamiltonian})
\cite{brehm99}.
Both Hamiltonians are identical except for couplings along the diagonals
if $J_\perp$ and $J_\parallel$ are suitably redefined \cite{notiz1}.
Here, we use Hamiltonian (\ref{eq:Hamiltonian})
since it is established that the four-spin terms are the significant ones
 \cite{mulle02}.
The exchange parameters determined in Ref.\ \onlinecite{nunne02} for
La$_{5.2}$Ca$_{8.8}$Cu$_{24}$O$_{41}$ correspond in our notation
\cite{notiz1} to
$J_\parallel/J_\perp = 1.22\pm 0.05$, $J_{\rm cyc}/J_\perp = 0.21\pm 0.03$
and $J_\perp = 1150 \pm 100$ cm$^{-1}$.
Using the average values, we find within 5\%  the same values
for the spin gap and for the two-triplet bound-state energies as in
Ref.\ \cite{nunne02}.

\begin{figure}[t]
\centerline{\psfig{figure=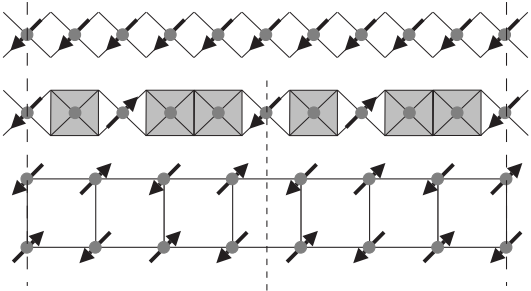,width=7cm,clip=}}
\caption{
Scheme\protect\cite{notiz2} 
of the superstructure along the chains and the ladders
($c$ axis). 10 chain units (top row) match 7 ladder units (bottom row)
inducing a modulation in the ladders with wave vector
$Q_{\rm S} = c_{\rm ladder}/c_{\rm chain} = \!10/7\!=\!3/7\!+\!1$
(in r.l.u.\ of the ladder) \protect\cite{frost97b}.
In Sr$_{14}$Cu$_{24}$O$_{41}$
the charge order (CO) implies an additional
superstructure with $Q_{\rm CO}= 2/10$
(in r.l.u.\ of the {\em chain})
\protect\cite{regna99b,matsu99,fukud02,brade02},
corresponding to a periodicity of $5 \! \cdot \! c_{\rm chain}$.
It is visualized (middle row) as two units of ``spin-hole-spin-hole-hole''
 per 7 rungs (grey squares denote the six holes per f.u.).
This superstructure induces
an additional modulation in the ladder with
$Q_{\rm CO} = c_{\rm ladder}/(5 \! \cdot \! c_{\rm chain}) = 2/7$
(in r.l.u.\ of the ladder).
\label{fig1}}
\end{figure}

The exchange coupling in the CuO$_2$ chains is much weaker than in
the ladders since it is mediated via Cu--O--Cu bonds with an angle
close to 90$^\circ$. Thus the chains do not contribute
directly to the Raman line at $\approx $ 3000 cm$^{-1}$. But the
presence of the chains gives rise to 7 inequivalent ladder rungs
per formula unit (f.u.) and thereby induces a modulation in the ladders
(see Fig.\ 1). This modulation is characterized by the wave vector
$Q_{\rm S}=10/7=3/7+1$ in reciprocal lattice units (r.l.u.) of the ladder.
In the magnetic subsystem of the spins on the Cu sites of the ladder,
wave vectors are only meaningful modulo unity so that $Q_{\rm S}=10/7$ and
$Q_{\rm S}=3/7$ are equivalent. The additional modulation induced by the
charge ordering on the chains below $T_{\rm CO} \approx 200$ K has the
wave vector $Q_{\rm CO} = 2/7$ \cite{regna99b,matsu99,fukud02,brade02},
see Fig.\ 1.

Now we estimate the amplitude of the exchange modulation with $Q_{\rm S}$.
The Cu-O distances within the ladder are hardly affected by the modulation;
the main effect is a shift of the O ions perpendicular to the Cu-O-Cu bonds
\cite{frost97b}. Hence the electronic hopping elements $t_{pd}$ can safely
be considered constant. The exchange couplings are modified by the induced
variation of the charge-transfer energy $\Delta_{\rm ct}$, i.e.\ the
variation of
the energy difference between holes on Cu and on O.\@ We computed the variation
of $\Delta_{\rm ct}$ in a point-charge model with stoichiometric valencies
except for the chain oxygen with $q \! = \! -1.7$e to account for the
holes \cite{nucke00,mizun97}.
The calculation uses Ewald sums so that the results pertain to the 
infinite system. The relative changes $\Delta J/J$ are estimated in leading
order \onlinecite{mulle02}.
Assuming structurally unmodulated chains and ladders,
 we find a negligible effect of the
chains on the exchange couplings of the ladder of $|\Delta J/J| \lessapprox 10^{-6}$.
However, the {\em modulated} positions at 300 K \cite{frost97b} yield
\begin{mathletters}
\label{roomtemp}
\begin{eqnarray}
J_{\parallel,i}  \!&=& \! J_\parallel[1+ 0.05
\cos(2\pi \textstyle\frac{3}{7}(i+\textstyle\frac{1}{2}))]\\
J_{\perp,i} \!&=& \! J_\perp[1 -0.10
\sin(2\pi \textstyle\frac{3}{7}i) +
0.05\cos(2\pi \textstyle\frac{6}{7}(i \!+ \!3))]
\end{eqnarray}
\end{mathletters}
with phase accuracy $|\Delta i| \lessapprox 0.1$ \cite{notiz2} where
 $i$ counts the leg- or the rung-bonds. The
term with $2Q_{\rm S} = 6/7$ denotes the second harmonic;
overtones with amplitude $\lessapprox$ 1\% are omitted. The
amplitudes in Eq.\ (\ref{roomtemp}) show that the induced modulation of
the couplings is indeed sizeable. 

We expect that the effects of the charge order occurring below
$T_{\rm CO} \approx 200$ K are of similar size. Without detailed
information on the structure at $T \ll T_{\rm CO}$, only an estimate is
possible. We assume a charge modulation on the chain oxygen of
$\Delta q(j) \! = \! - 0.2 \mbox{e}\, \cos(2\pi \frac{2}{10}
 (j+\frac{1}{2}))$,
where $j$ counts the chain O sites, and the periodicity $5c_{\rm chain}$
and the phase are established experimentally
\cite{regna99b,matsu99,fukud02,brade02} (cf.\ middle row of Fig.\ 1).
This yields an additional modulation
\begin{equation}
\label{modCO}
\Delta J_{\parallel,i}= 0.16 J_\parallel\cos(2\pi \textstyle\frac{2}{7}i)\ ,
\end{equation}
with phase accuracy of $|\Delta i| \lessapprox 0.1$ \cite{notiz2}.
Thus the modulation induced by the charge order with $Q_{\rm CO}= 2/7$
(in r.l.u.\ of the ladder) is indeed significant.

Now, we investigate the effects of modulations on magnetic Raman
 scattering which probes the excitations with zero momentum
and zero spin. 
At $T$=0 the Raman response $I(\omega)$ is 
given by the retarded resolvent
\begin{equation}
\label{Intensity}
I(\omega) = {\textstyle\frac{-1}{\pi}}
\lim_{\delta\to0+} \textstyle{\rm Im}\left\langle0\left|
R^{\dagger}\frac{1}{\omega-H+E_0+i\delta}R\right|0\right\rangle .
\end{equation}
The observables $R^{\rm rung}$ ($R^{\rm leg}$) for magnetic light scattering in
rung-rung (leg-leg) polarization are given in Ref.\ \cite{schmi01}. We focus on
the dominant two-triplet contribution. A CUT is employed to map the Hamiltonian
$H$ to an effective Hamiltonian $H_{\rm eff}$ which conserves the number of
rung-triplets \cite{knett00a,knett01b,schmi01}. The ground state of $H_{\rm eff}$
is the rung-triplet vacuum. The observable
$R$ in $I(\omega)$
is mapped by the same unitary transformation to an effective observable 
$R_{\rm eff}$.

The CUT is implemented perturbatively in $J_\parallel/J_{\perp}$. We compute
$H_{\rm eff}$ and $R_{\rm eff}$ to order $n \geq 10$. A calculation in order
$n$ accounts for hopping and interaction processes extending over a distance
of $n$ rungs. The resulting plain series are represented in terms of the 
variable $1-\Delta/(J_\parallel+J_\perp)$ \cite{schmi02a}. Then standard 
Pad\'e approximants
\cite{domb89} yield reliable results up to $J_\parallel/J_{\perp}\approx 1-1.5$
depending on the value of $J_{\rm cyc}/J_{\perp}$.
Consistency checks were carried out by extrapolating the involved quantities
before and after Fourier transforms. In case of inconclusive approximants the
bare truncated series are used. We estimate the overall accuracy to be 
$\approx $5\%. The Raman line shape is finally calculated as
continued fraction by tridiagonalization of the effective two-triplet Hamiltonian
in a mixed representation using the total momentum and the real-space distance.
So the total momentum is sharply defined. No finite-size effects appear. This
ensures a particularly high resolution in momentum and in energy necessary
to account for a very sharp feature.

The modulation is included on the level of the effective model, i.e.\
\emph{after} the CUT.\@ This is no serious caveat since
a microscopic calculation is not available.
The leading-order effect of $J_\parallel$ is to enable the elementary triplets
to hop from rung to rung by a nearest-neighbor hopping element
$t_1 \propto J_\parallel$ and to induce a nearest-neighbor interaction
$w_1 \propto J_\parallel$.
So the most straightforward way to account for the
modulation of $J_\parallel$ as given in Eqs.\ (\ref{roomtemp}a) and
(\ref{modCO}) is to modulate $t_1$ and $w_1$,
\begin{equation}
\label{eq:mod}
t_1 \propto w_1 \propto J_\parallel \cdot [ \;
1 + \sum_{Q=2/7,\; 3/7,\; 6/7} \alpha_{Q} \cos(2\pi\textstyle Q i)
\;] \; .
\end{equation}
Since we focus on the effect of the charge order (Eq.\ \ref{modCO}), the
modulation of $J_\perp$ as given in Eq.\ (\ref{roomtemp}b) is neglected.

We use the parameters fixed for La$_{5.2}$Ca$_{8.8}$Cu$_{24}$O$_{41}$
in Refs.\ \onlinecite{nunne02,notiz1}, $J_\parallel/J_\perp$=1.25 and
$J_{\rm cyc}/J_\perp$=0.2.
Fig.\ 2 shows the dispersion with and without a 15\% modulation with
wave vector $Q_{\rm CO}$, i.e.\ $\alpha_{2/7}$=0.15 in Eq.\ (\ref{eq:mod}).
Clearly, sizeable gaps open wherever $Q_{\rm CO}$ links equal energies
$\omega(k)=\omega(k \! + \! Q_{\rm CO})$ of the unmodulated ladder.
Smaller gaps open for higher-order processes, e.g.\ for
$\omega(k)=\omega(k \! + \! 2Q_{\rm CO})$. Thus the energies at which gaps
open depend decisively on the wave vector of the modulation.

\begin{figure}[t]
\centerline{\psfig{figure=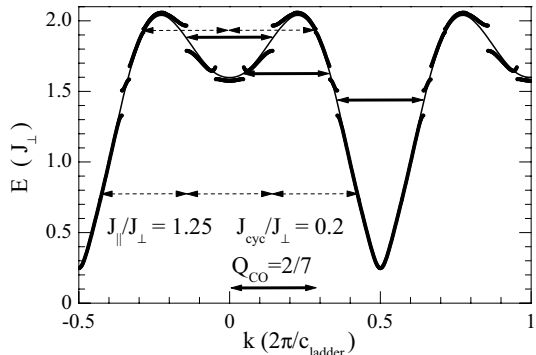,width=7cm,clip=}}
\caption{Dispersion of the elementary triplets
with (thick line) and without (thin line) a modulation of
$\alpha_{2/7}$=0.15 (cf.\ Eq.~\ref{eq:mod}) with $Q_{\rm CO}=2/7$
(arrows). Some higher-order contributions are denoted by dashed arrows.
}
\label{fig2}
\end{figure}

\begin{figure}[t]
\centerline{\psfig{figure=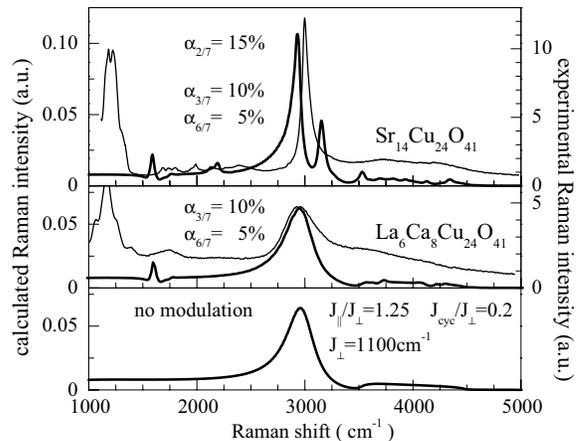,width=7.5cm,clip=}}
\caption{
Thick lines: Raman line shapes in leg-leg polarization for
$J_\parallel/J_\perp \! = \! 1.25$, $J_{\rm cyc}/J_\perp \! = \! 0.2$
and $J_\perp$=1100 cm$^{-1}$ \protect\cite{nunne02,notiz1} without
modulation (bottom), with the structural modulation $Q_S$=3/7 and 6/7
appropriate for La$_{6}$Ca$_{8}$Cu$_{24}$O$_{41}$ (middle) and with
the additional charge-order modulation of Sr$_{14}$Cu$_{24}$O$_{41}$ (top).
Thin lines: Raman data from Ref.\ \protect\cite{sugai99}, T=20K.\@
A modulation-induced gap in the dispersion at $\omega_g$ (see Fig.\ 2)
causes a Raman feature at 2$\omega_g$.
Additional features arise due to backfolding, e.g.\ the small peak at
2200 cm$^{-1}$ corresponds to the $S$=0 two-triplet bound state at
$k$=2/7 \protect\cite{knett01b}.
}
\label{fig3}
\end{figure}

The Raman response of an undoped and unmodulated ladder is very broad (bottom
panel of Fig.\ 3 and Ref.\ \cite{schmi01}), in good agreement with data of
La$_{6}$Ca$_{8}$Cu$_{24}$O$_{41}$ \cite{sugai99} (middle panel of Fig.\ 3).
The excellent description of the peak position obtained for the parameter set
 of Ref.\ \onlinecite{nunne02} given above  corroborates these parameters.
What is the effect of a modulation on the Raman line shape?
The occurrence of gaps implies prominent peaks (van Hove singularities) in
the density of states (DOS) and hence in the Raman line shape. Since Raman
scattering measures excitations with total momentum $k_{\rm tot}$=0, the
two-triplet response reflects the excitation of two triplets with momenta
$k_2 \! = \! - k_1$ and  energies $\omega(k_1) \! = \! \omega(k_2)$. A gap
at $\omega_g$ thus causes a corresponding feature in the Raman line at 
$2\omega_g$. For the structural wave vectors $Q_{\rm S}$=3/7 and 6/7 these
effects are rather small (middle panel of Fig.\ 3). But a drastic change of the
line shape appears if (and only if) $2\omega_g$ coincides with the 
broad peak of the unmodulated ladder, since then the opening of the gap
implies a redistribution of  a large part of the spectral weight.
 For the relevant exchange couplings we find that
$2 \cdot \omega(\pi/c_{\rm ladder} \pm Q_{\rm CO}/2) \approx 2.8 J_\perp 
\approx 3100$ cm$^{-1}$
is slightly above the Raman peak of the unmodulated ladder. Hence the
charge-order modulation piles up a large part of the high-frequency weight
on top of the peak, giving rise to a very sharp feature
 which agrees very well with the data of Sr$_{14}$Cu$_{24}$O$_{41}$ \cite{sugai99,gozar01}
(top panel of Fig.\ 3).
Since both the exchange constants in Sr$_{14}$Cu$_{24}$O$_{41}$ and the calculations
are accurate  within a few percent, only a semi-quantitative analysis 
is possible which shows the principal mechanism. The remaining uncertainties
may imply that also a smaller
 value of $\alpha_{2/7}$ is sufficient to produce the sharp feature.

The good agreement between the experimental data and the theoretical result,
based on the independently determined couplings \cite{nunne02,notiz1} and the
wave vector of the charge order,  corroborates our interpretation.
Another argument stems from the polarization dependence.
For $J_{\rm cyc}>0$, the peak positions for leg-leg and rung-rung polarization
should be different \cite{schmi01}. This is indeed the case in
La$_{6}$Ca$_{8}$Cu$_{24}$O$_{41}$ \cite{sugai99}, but
the sharp peak in Sr$_{14}$Cu$_{24}$O$_{41}$ is found at
$\approx$ 3000 cm$^{-1}$ in {\em both} polarizations \cite{sugai99,gozar01}.
In the scenario of the modulation-induced gaps the peak position is determined
by the position and the size of the gap, since the peak is primarily a DOS
effect. Hence the coincidence of the peak positions in both polarizations
despite $J_{\rm cyc}>0$ supports our scenario. Moreover, also the spectra
of Sr$_{14}$Cu$_{24}$O$_{41}$ at elevated temperatures are explained.
The charge order melts at $T_{\rm CO} \approx $ 200 K. Indeed, the very sharp
Raman line is observed only below $T_{\rm CO}$ \cite{sugai99}.
For $T\geq T_{\rm CO}$, the peak positions are {\em different} for the two
polarizations \cite{sugai99}, which is expected for
$J_{\rm cyc}\approx 0.2J_\perp$ at $\alpha_{2/7}=0$.

An alternative explanation of the sharp peak 
in terms of bound states is unlikely.
There is no bound state within the broad
Raman continuum of the undistorted, undoped ladder \cite{schmi01}.
But how about finite doping? At 300 K, about 90 \% of the doped carriers
reside in the chains \cite{nucke00}. At low temperatures the
distribution of holes is not yet settled experimentally. Theoretically,
the Madelung potentials indicate that all the holes reside in
the chains \cite{mizun97}. This is corroborated by the observation
of the periodicity $5c_{\rm chain}$ \cite{regna99b,matsu99,fukud02,brade02}
for the charge order in the chains. In a 1D fermionic system it is natural
to view the charge order
as an effect of the $2k_{\rm F}$ instability. So we are led to conclude that
$2k_{\rm F}= 2/10$ (in r.l.u.\ of the chain),
which implies that there are
$n_\downarrow+n_\uparrow = 4/10$
electrons per site or 6 holes per f.u.. This further supports our assumption
that at $T\approx 0$ all holes reside on the chains.

A finite hole concentration on the ladders cannot be ruled out completely.
These charges would be pinned in a commensurate charge-density wave (CDW)
at low temperatures by the electrostatic potential of the charge order on
the chains. Clearly, such a CDW would also induce strong modulations.
But it remains unclear where the peculiar periodicity stems from if
$2k_{\rm F}\neq 2/10$.

A small amount of impurity holes cannot explain the sharp Raman peak.
Below $T_{\rm CO}$, Sr$_{14}$Cu$_{24}$O$_{41}$
is insulating, i.e.\ all charge carriers are localized. The local
charge degrees of freedom may couple to the magnetic ones, but due to
the local character the whole Brillouin zone would be involved implying a
\emph{broad} energy distribution, at odds with experiment \cite{sugai99,gozar01}.

Our results clearly call for several experimental verifications.
Neutron-scattering experiments could clarify the presence and the size of
gaps in the dispersion. Low-temperature investigations of the structure would
help to improve our understanding of the charge-ordered state.
Low-temperature x-ray absorption data are required to determine the hole
density in the ladders. Raman studies as a function of Ca
concentration and temperature could verify that the features explained here
are indeed connected to the occurrence of the charge-ordered
state. Then, the peak position 
offers a sensitive tool to determine
the modulation wave vector $Q_{\rm CO}$.

In conclusion, the modulation of the exchange coupling in the charge-ordered
state of Sr$_{14}$Cu$_{24}$O$_{41}$ can explain the peculiar Raman data.
The induced gap redistributes a large part of the spectral weight,
giving rise to a sharp Raman peak.
A comparison with the 2D cuprates is not appropriate.
Strong quantum fluctuations are still the most likely candidate to explain their
very broad Raman line.

We thank E. M\"{u}ller-Hartmann, G. Blumberg, A. Gozar, and
M. Braden for helpful
discussions and the DFG for financial support in SP 1073 and in SFB 608.

%\bibliographystyle{prsty}
%\bibliography{../bibinput/liter10}

\begin{thebibliography}{10}
\bibitem{notiz1}
Neglecting the two-spin exchange terms along the diagonals,
our notation
$J_\perp, J_\parallel, J_{\rm cyc}$ is given in terms of the other
notation $J^{\rm p}_\perp, J^{\rm p}_\parallel, J^{\rm p}_{\rm
cyc}$ using cyclic permutations by $J_\parallel =J^{\rm
p}_\parallel +\frac{1}{4}J^{\rm p}_{\rm cyc}$, $J_\perp =J^{\rm
p}_\perp +\frac{1}{2}J^{\rm p}_{\rm cyc}$, and $J_{\rm cyc}=J^{\rm
p}_{\rm cyc}$ \protect\cite{brehm99}.
\bibitem{notiz2}
Fig.~1 shows the periodicity; the
 modulation {\em phases} of the couplings cannot be read off
 because (i) Cu sites in the chains and ladders are not
 in line; (ii) each ladder  is influenced by the four 
chains close by which are shifted by different amounts;
(iii) the important oxygens are not depicted.
 For details see Ref.~\protect\onlinecite{frost97b}.
\end{thebibliography}

\end{document}